\documentclass[10pt,a4paper,twocolumn]{article}
\usepackage[utf8]{inputenc}

\usepackage[english]{babel}

\usepackage{hyperref}
\usepackage{soul}
\usepackage{xcolor}
\usepackage{url}
\usepackage{multirow}
\usepackage{subcaption}
\usepackage{graphicx}
\usepackage{url}
\usepackage[numbers,compress]{natbib}
\usepackage{hyperref}

\usepackage[margin=0.75in]{geometry}

\begin{document}



\title{\textbf{A Reflection on the Organic Growth of the Internet Protocol Stack}}

\author{Jordi Paillisse$^{\dag}$, Alberto Rodriguez-Natal$^{\ddag}$, Fabio Maino$^{\ddag}$ and  Albert Cabellos$^{\dag}$ \\
$^{\dag}$\textit{\small UPC-BarcelonaTech, Barcelona, Spain} - \textit{\small \{jordi.paillisse, alberto.cabellos\}@upc.edu,} \\
$^{\ddag}$\textit{\small Cisco, San Jose, CA, USA - \{natal, fmaino\}@cisco.com}}

\maketitle

\begin{abstract}

In the last 15 years, the Internet architecture has continued evolving organically, introducing new headers and protocols to the classic TCP/IP stack. More specifically, we have identified two major trends. First, it is common that most communications are encrypted, either at L3 or L4. And second, due to protocol ossification, developers have resorted to upper layers to introduce new functionalities (L4 and above). For example, QUIC's connection migration feature provides mobility at L4. 

In this paper we present a reflection  around these changes, and attempt to formalize them by adding two additional protocol headers to the TCP/IP stack: one for security, and another for new functionalities. We must note that we are not presenting a new architecture, but trying to draw up what it's already out there. 
In addition, we elaborate on the forces that have brought us here, and we enumerate current proposals that are shaping these new headers. We also analyze in detail three examples of such trends: the Zero Trust Networking paradigm, the QUIC transport protocol, and modern SD-WAN systems. Finally, we present a formalization of this architecture by adding these two additional layers to the TCP/IP protocol stack. Our goal is triggering a discussion on the changes of the current Internet architecture.

\end{abstract}

\section{Introduction}

The changes in the Internet in the last 15 years have outlined a different architecture than that of TCP/IP. Two key trends have driven these changes:  traffic encryption, and protocol ossification.

First, due to the increasing encryption of all Internet traffic (about 70\% of  traffic on the Internet is encrypted \cite{trafficEnryptedInternet, rfc7258PervasiveMonitoring}), it is becoming more and more difficult to add advanced network functionalities to the Internet protocol stack. In turn, this has   increased the complexity of performing common network functions such as traffic classification or Deep Packet Inspection \cite{dpiEncryptedTraffic, encryptedTrafficClassSurvey}.

Second, protocol ossification has forced developers and vendors to move innovation to upper layers in the stack in order to be able to introduce new functionalities. This has introduced new headers and layers in the stack. QUIC is the most prominent example, because it uses UDP to cross the Internet. Or the Zero Trust paradigm, which is mainly based in HTTPS. This has already been outlined in multiple publications that discuss how to avoid Internet ossification and deploy new services \cite{extensibleInternet, permanentRevolutionInternetArch, evolvableInternetShenker05}, or stress the importance of HTTP as the new narrow waist of the Internet \cite{httpNarrowWaist}.

The consequence of these two driving forces is that we can observe two new layers added to the classical TCP/IP stack. First, a security layer, usually defined by standards like IPsec or TLS. Second, an extension layer that carries information for the aforementioned additional functionality, which is not necessarily standardized and depends on each vendor or application requirements.

In this paper we present an attempt to describe such architecture. Note that we are not proposing a new architecture, rather describing something that's already present on the current Internet. Our goal is triggering a discussion on the changes of the current Internet architecture. First, we enumerate and describe current  proposals that are shaping these new headers. Then we briefly analyze three of these proposals, namely Zero Trust Networking, SD-WANs (Software-Defined Wide Area Networks), and the QUIC (Quick UDP Internet Connections) transport protocol, and discuss how they make use of the two additional headers. Finally, we formalize this rearranging of the Internet stack,  describing the functions of each layer.

\begin{figure*}[!htb]
\centering
\includegraphics[width = \textwidth, keepaspectratio  ,trim={0cm, 0cm, 0cm, 0cm}, clip]{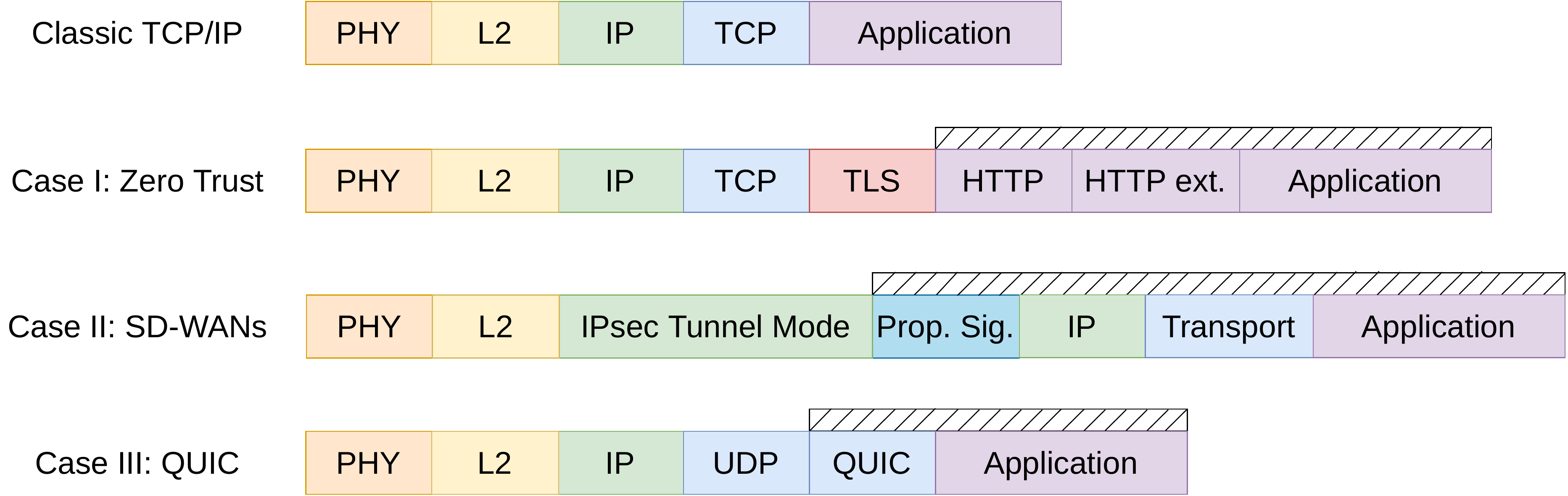}
\caption{Comparison of headers from the classical TCP/IP stack with the three use cases. The striped box represents  an encrypted header. \textit{Prop. Sig.} stands for the proprietary signaling of some SD-WAN vendors.}
\label{fig:headers}
\end{figure*}

\section{Current Trends}
\label{sec:trends}

As we mentioned, we can observe the emergence of two additional headers in the classic Internet stack, one to encrypt communications (Security header), and another to add new functionalities (Extension header). In this section we briefly describe several examples of them. Note that this list does not pretend to be exhaustive, but rather a quick summary of current and emerging proposals.

\subsection{Security header}

\begin{description}
\item[Classical protocols:] IPsec \cite{rfc4301ipsec} and Transport Layer Security \cite{tls1-3} are the two paramount examples of a header that offers data confidentiality, integrity and/or authentication. 

\item[WireGuard:] WireGuard \cite{donenfeld2017wireguard} is an emerging VPN protocol that uses a UDP header like QUIC, and encapsulates traffic similarly to the IPsec tunnel mode. Its main features are simple configuration with SSH-like keys, NAT traversal,  mobility, and a fast implementation in the Linux kernel .

\item[Zero Trust Networks:] the Zero Trust paradigm uses TLS as the security layer. The key difference with classical TLS is the usage of the \emph{access proxy}, that terminates client TLS connections, verifies if the connection is authorized, and creates a new TLS connection towards the destination server. Section \ref{sec:zero_trust} discusses the architectural implications of this paradigm.

\item[Service Meshes:] service meshes \cite{serviceMeshSoA} also leverage TLS to encrypt communications, and operate similarly to Zero Trust Networks. Instead of the access proxy, they use a \emph{sidecar proxy} to intercept all communications between applications, monitor their status, and enforce policies. Some of them use mutual TLS to authenticate both ends of the connection. Although they are currently restricted to the data center, this could change with Envoy mobile \cite{envoyMobile}, since it can be deployed in end-hosts.

\end{description}

\subsection{Extension header}

\begin{description}
\item[Network Tokens:] Network Tokens \cite{network-tokens} proposes including small cryptographic signatures in the extension headers of existing protocols. These tokens carry variable information such as SLA requirements, or the type of application, and they can be used to perform traffic differentiation in service provider networks.
\item[Application-aware Networking :] the Application-aware Networking (APN \cite{apn-problem-statement}) proposal in the IETF is in the process of defining a label for fine-grain QoE, that includes application and user identification \cite{apnHeaderFields}. APN targets several use cases for networks in single administrative domains, such as 5G, network slicing, online gaming, SD-WANs, etc.
\item[MASQUE:] the IETF MASQUE (Multiplexed Application Substrate over QUIC Encryption) Working Group \cite{masque-protocol} is creating a framework to multiplex several application flows on top of QUIC, and to proxy all kinds of IP traffic over QUIC and HTTP/3. Some use cases are HTTP proxies, DNS over HTTP, QUIC proxying, or forwarding of UDP frames. 
\item[eBPF:] the extended Berkeley Packet Filter (eBPF) allows programming the Linux kernel network stack via a JIT compiler and a virtual machine.  Although eBPF is not a protocol per se, this programmability can be used to create new protocols, or to extend existing ones, such as TCP \cite{ebpfTCP}. 
\end{description}

\begin{figure*}[!htb]
\centering
\includegraphics[width =0.8 \textwidth, keepaspectratio  ,trim={0cm, 0cm, 0cm, 0cm}, clip]{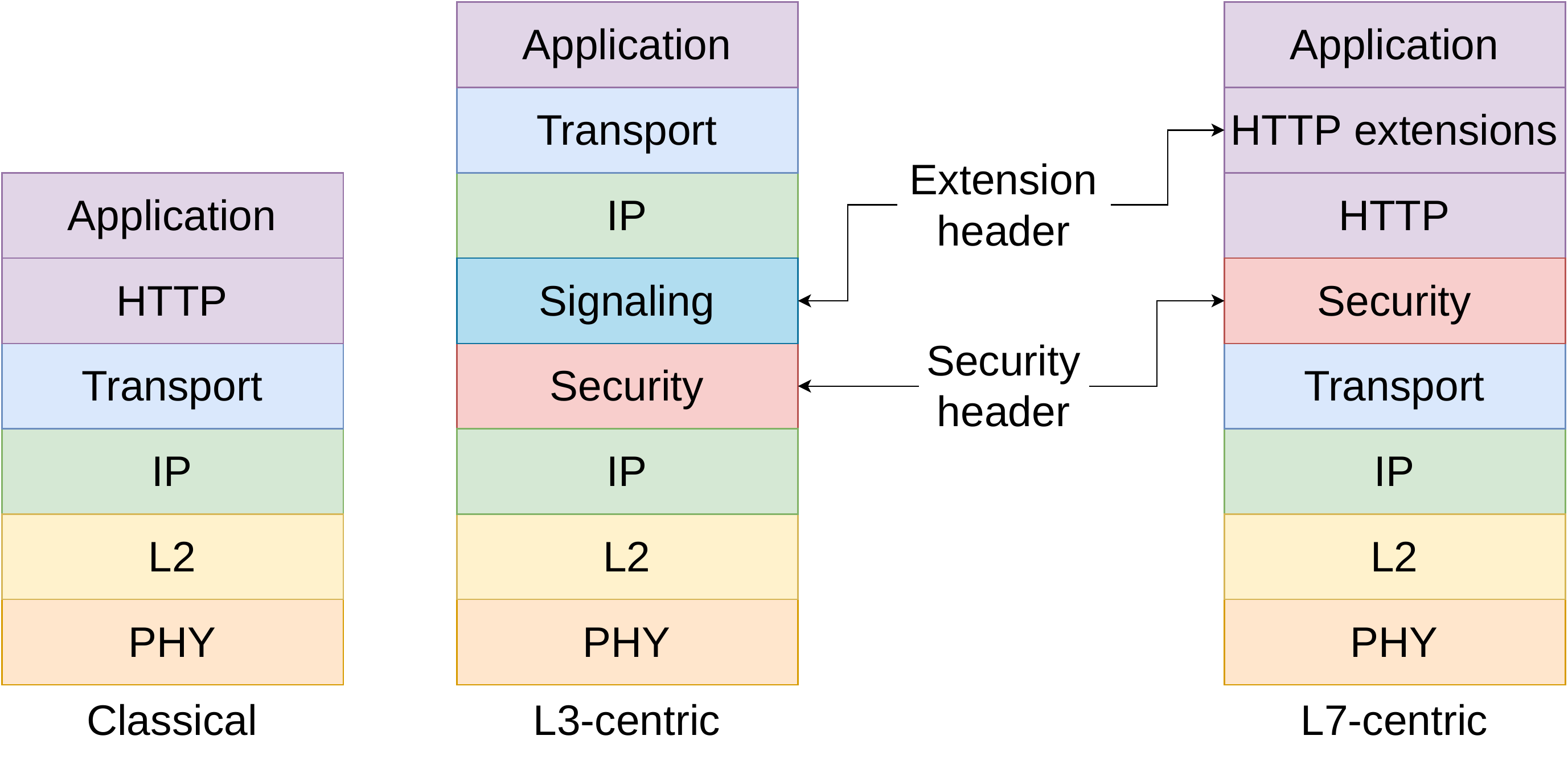}
\caption{Classical Internet stack (left), and proposed stacks: L3-centric (middle) and L7-centric (right).}
\label{fig:layers}
\end{figure*}

\section{Case Studies}

This section describes how three recent network paradigms and/or protocols, namely Zero Trust Networks, SD-WANs, and QUIC leverage the Security and Extension headers.

\subsection{Zero Trust Networks} 
\label{sec:zero_trust}

The security header in the  Zero Trust paradigm \cite{beyondCorpGoogleAccessProxy, oreillyZeroTrust} corresponds to TLS, since all communications leverage in HTTPS (second row of fig. \ref{fig:headers}). With respect to the extension header, this paradigm makes use of well known authentication options (Open ID Connect, OAuth) to identify users and devices, as well as custom extra HTTP headers to identify and classify traffic flows and applications \cite{beyondCorpGoogleAccessProxy}.  Then, this classification capability is combined with complex policies, such as preventing access to an application to devices running an outdated OS. This contrasts with L3 protocols in two aspects. First, security headers in L3 usually offer confidentiality and authentication, but not user identity (the notable exception are IP-based ACLs). Second, it is more difficult to implement complex policies at L3, which commonly use labels, like the IPv6 Flow Label or MPLS.

\subsection{SD-WANs}
Commercial SD-WANs are ubiquitous nowadays \cite{sdwanInetCensus}, and we can also find both the security and the extension headers in their design. In this case, the security header appears often in the form of IPsec in tunnel mode, or equivalent L3 security protocols. Regarding the extension header, it is common that  these systems add an additional header right after the security header (Prop. Sig. in fig. \ref{fig:headers}, third row). This header can be used for different purposes, the most common being isolation, i.e. creating isolated VPNs for security reasons. Other use cases are bandwidth aggregation, redundancy, multicast, etc. A classical example of this header is VXLAN \cite{rfc7348vxlan}, or  Multiprotocol Label Switching (MPLS) labels.

\subsection{QUIC}
\label{sec:quic}
Our last example is the QUIC transport protocol \cite{rfc9000quic}. We  consider the QUIC header as an extension header, because it is right after a UDP header (fig. \ref{fig:headers}, bottom row). QUIC makes use of this header to provide classic transport functions (connection establishment, congestion control, etc), but it is also used to provide additional features. The two most relevant are the connection migration feature, that implements mobility at L4, and the multipath extensions \cite{deconinck-quic-multipath-07}. Finally, note that QUIC adds the security header in the form of TLS \cite{rfc9001quicTLS}.

\section{Architecture}

Finally, taking into account the trends from sec. \ref{sec:trends}, and the three case studies we have discussed, we present a formalization in two different stacks (fig. \ref{fig:layers}): one that mainly leverages L3 headers (L3-centric), and another one that adds most functionality at HTTP level (L7-centric). In what follows, we describe the functions of each layer and differences between the two approaches, when applicable.

\textbf{PHY and L2:} physical and link layer protocols, respectively.

\textbf{Routing:} a plain IP header, able to traverse the public Internet. It can include a UDP header to accommodate new protocols, eg. QUIC or WireGuard.

\textbf{Security:} a header with security information to provide data confidentiality and integrity. We can find it right after the IP header in the L3-centric approach (e.g. IPsec, WireGuard), or as a TLS header after the transport header in the case of the L7-centric stack. Note that these two headers are not -generally speaking- interchangeable because they were designed with different objectives in mind: point to point vs. end-to-end encryption, respectively. However, it appears that some paradigms are challenging this assumption, such as Zero Trust Networks, that concatenate two or more TLS tunnels with HTTPS proxies in between.

\textbf{Extension header:} This header provides additional functionalities. In the L3-centric approach (middle of fig. \ref{fig:layers}), we can find it right after the security header. The most common example is the signaling header in SD-WANs that contains the VPN identifier to provide network isolation, group tags, etc. In the L7 approach, this header appears in the form of HTTP extensions, e.g. in Zero Trust networks it carries user identification. We must remark that this header is not necessarily standardized, as opposed to the security header. It can carry information for multiple purposes, like identity, QoE, policy labels, etc.

\textbf{Transport:} common transport protocols, like TCP, UDP, or more recently QUIC.

\textbf{HTTP and application:} the common HTTP headers and the application data.

\begin{table*}[!htb]
\caption{Feature comparison between L3 and L7-centric approaches}
\begin{center}
\begin{tabular}{|l|c|c|}
\hline
\textbf{Feature}  &  \textbf{L3-centric}  & \textbf{L7-centric}  \\
\hline

Routing         & IP-based  & URL-based  \\
\hline
Load balancing  &  Simple: 5 tuple or ECMP &   Rich, eg. service meshes   \\
\hline
Application awareness & DPI, port and protocol number & Native, as complex as desired  \\
\hline
\multirow{2}{*}{Feature velocity}    &  Slow in ASICs      &      \multirow{2}{*}{High}     \\
                                     &   High in software       &          \\
   
\hline
Policy & MPLS, tagging, IP ACLs &  Rich, policy databases \\
\hline
\multirow{2}{*}{Performance}       & High in ASICs         &  Medium-low \\
                                    & Medium in software   &   (software) \\
\hline
Process encrypted traffic   &  Only L3-encrypted & Yes   \\
\hline
\end{tabular}
\label{tab:comparison}
\end{center}
\vspace{-0.55cm}
\end{table*}

Taking into consideration the two proposed stacks, table \ref{tab:comparison} presents a comparison of several features between the L3-centric approach and the L7 one. Some of them are worth mentioning, such as load balancing: while at L3 we leverage 5-tuple or Equal Cost Multi-Path (ECMP), at L7 we can implement complex load balancing strategies, like in service meshes, that allow A/B testing or canary releases. On the same line, making the network aware of the application is simple at L7, because we actually are at the application layer. On the contrary, at L3 we must rely on DPI and port and protocol numbers. The latter are not enough in some situations, since a significant part of applications run on top of HTTP(S). 

On the performance side, it is common to implement L3 network functions in ASICs, which yields high performance implementations, while most L7 networking functions are implemented in software. This situation may change due to the introduction of programmable switches based on P4 \cite{p4paper}. Finally, a L7 approach can process all kinds of encrypted traffic\footnote{Some Zero Trust architectures encapsulate traffic from lower layers into HTTP headers to be able to process it.}, while L3 is limited to IPsec traffic and similar protocols.

\section{Conclusion}
In this paper we have examined some proposals shaping the present and the future of Internet protocols. We have suggested adding two additional headers to the classical TCP/IP stack, in order to accommodate the evolution of the Internet architecture in the last 15 years. First, the security header, that provides confidentiality, authentication and data integrity. Second, the extension header, that supports additional functionalities. We have also examined three use cases that make use of these headers. Finally, we have presented a formalization of this stack in two possible stacks: a L3-centric, and a L7-centric.

\bibliographystyle{unsrtnat}
\begin{small}
\bibliography{layers_paper_bibliography, additional_bib}

\begin{thebibliography}{26}
\providecommand{\natexlab}[1]{#1}
\providecommand{\url}[1]{\texttt{#1}}
\expandafter\ifx\csname urlstyle\endcsname\relax
  \providecommand{\doi}[1]{doi: #1}\else
  \providecommand{\doi}{doi: \begingroup \urlstyle{rm}\Url}\fi

\bibitem[Maddison(2018)]{trafficEnryptedInternet}
John Maddison.
\newblock Encrypted traffic reaches a new threshold, nov 2018.
\newblock URL
  \url{https://www.networkcomputing.com/network-security/encrypted-traffic-reaches-new-threshold}.

\bibitem[Farrell and Tschofenig(2014)]{rfc7258PervasiveMonitoring}
Stephen Farrell and Hannes Tschofenig.
\newblock {Pervasive Monitoring Is an Attack}.
\newblock RFC 7258, May 2014.
\newblock URL \url{https://rfc-editor.org/rfc/rfc7258.txt}.

\bibitem[Sherry et~al.(2015)Sherry, Lan, Popa, and
  Ratnasamy]{dpiEncryptedTraffic}
Justine Sherry, Chang Lan, Raluca~Ada Popa, and Sylvia Ratnasamy.
\newblock Blindbox: Deep packet inspection over encrypted traffic.
\newblock In \emph{Proceedings of the 2015 ACM Conference on Special Interest
  Group on Data Communication}, SIGCOMM '15, page 213–226, New York, NY, USA,
  2015. Association for Computing Machinery.
\newblock ISBN 9781450335423.
\newblock \doi{10.1145/2785956.2787502}.
\newblock URL \url{https://doi.org/10.1145/2785956.2787502}.

\bibitem[Cao et~al.(2014)Cao, Xiong, Zhao, Li, and
  Guo]{encryptedTrafficClassSurvey}
Zigang Cao, Gang Xiong, Yong Zhao, Zhenzhen Li, and Li~Guo.
\newblock A survey on encrypted traffic classification.
\newblock In Lynn Batten, Gang Li, Wenjia Niu, and Matthew Warren, editors,
  \emph{Applications and Techniques in Information Security}, pages 73--81,
  Berlin, Heidelberg, 2014. Springer Berlin Heidelberg.
\newblock ISBN 978-3-662-45670-5.
\newblock \doi{10.1007/978-3-662-45670-5_8}.

\bibitem[Balakrishnan et~al.(2021)Balakrishnan, Banerjee, Cidon, Culler,
  Estrin, Katz-Bassett, Krishnamurthy, McCauley, McKeown, Panda, Ratnasamy,
  Rexford, Schapira, Shenker, Stoica, Tennenhouse, Vahdat, and
  Zegura]{extensibleInternet}
Hari Balakrishnan, Sujata Banerjee, Israel Cidon, David Culler, Deborah Estrin,
  Ethan Katz-Bassett, Arvind Krishnamurthy, Murphy McCauley, Nick McKeown,
  Aurojit Panda, Sylvia Ratnasamy, Jennifer Rexford, Michael Schapira, Scott
  Shenker, Ion Stoica, David Tennenhouse, Amin Vahdat, and Ellen Zegura.
\newblock Revitalizing the public internet by making it extensible.
\newblock \emph{SIGCOMM Comput. Commun. Rev.}, 51\penalty0 (2):\penalty0
  18–24, May 2021.
\newblock ISSN 0146-4833.
\newblock \doi{10.1145/3464994.3464998}.
\newblock URL \url{https://doi.org/10.1145/3464994.3464998}.

\bibitem[McCauley et~al.(2019)McCauley, Harchol, Panda, Raghavan, and
  Shenker]{permanentRevolutionInternetArch}
James McCauley, Yotam Harchol, Aurojit Panda, Barath Raghavan, and Scott
  Shenker.
\newblock Enabling a permanent revolution in internet architecture.
\newblock In \emph{Proceedings of the ACM Special Interest Group on Data
  Communication}, SIGCOMM '19, page 1–14, New York, NY, USA, 2019.
  Association for Computing Machinery.
\newblock ISBN 9781450359566.
\newblock \doi{10.1145/3341302.3342075}.
\newblock URL \url{https://doi.org/10.1145/3341302.3342075}.

\bibitem[Ratnasamy et~al.(2005)Ratnasamy, Shenker, and
  McCanne]{evolvableInternetShenker05}
Sylvia Ratnasamy, Scott Shenker, and Steven McCanne.
\newblock Towards an evolvable internet architecture.
\newblock \emph{SIGCOMM Comput. Commun. Rev.}, 35\penalty0 (4):\penalty0
  313–324, August 2005.
\newblock ISSN 0146-4833.
\newblock \doi{10.1145/1090191.1080128}.
\newblock URL \url{https://doi.org/10.1145/1090191.1080128}.

\bibitem[Popa et~al.(2010)Popa, Ghodsi, and Stoica]{httpNarrowWaist}
Lucian Popa, Ali Ghodsi, and Ion Stoica.
\newblock Http as the narrow waist of the future internet.
\newblock In \emph{Proceedings of the 9th ACM SIGCOMM Workshop on Hot Topics in
  Networks}, Hotnets-IX, New York, NY, USA, 2010. Association for Computing
  Machinery.
\newblock ISBN 9781450304092.
\newblock \doi{10.1145/1868447.1868453}.
\newblock URL \url{https://doi.org/10.1145/1868447.1868453}.

\bibitem[Seo and Kent(2005)]{rfc4301ipsec}
Karen Seo and Stephen Kent.
\newblock {Security Architecture for the Internet Protocol}.
\newblock RFC 4301, December 2005.
\newblock URL \url{https://rfc-editor.org/rfc/rfc4301.txt}.

\bibitem[Rescorla(2021)]{tls1-3}
Eric Rescorla.
\newblock {The Transport Layer Security (TLS) Protocol Version 1.3}.
\newblock Internet-Draft draft-ietf-tls-rfc8446bis-01, Internet Engineering
  Task Force, February 2021.
\newblock URL
  \url{https://datatracker.ietf.org/doc/html/draft-ietf-tls-rfc8446bis-01}.
\newblock Work in Progress.

\bibitem[Donenfeld(2017)]{donenfeld2017wireguard}
Jason~A Donenfeld.
\newblock Wireguard: Next generation kernel network tunnel.
\newblock In \emph{NDSS}, 2017.
\newblock \doi{10.14722/ndss.2017.23160}.
\newblock URL \url{http://dx.doi.org/10.14722/ndss.2017.23160}.

\bibitem[Li and Lemieux~et al(2019)]{serviceMeshSoA}
Wubin Li and Yves Lemieux~et al.
\newblock Service mesh: Challenges, state of the art, and future research
  opportunities.
\newblock In \emph{2019 IEEE International Conference on Service-Oriented
  System Engineering (SOSE)}, pages 122--1225, 2019.
\newblock \doi{10.1109/SOSE.2019.00026}.

\bibitem[Project(2021)]{envoyMobile}
Envoy Project.
\newblock Envoy mobile, july 2021.
\newblock URL \url{https://envoy-mobile.github.io/}.

\bibitem[Yiakoumis et~al.(2020)Yiakoumis, McKeown, and
  Sorensen]{network-tokens}
Yiannis Yiakoumis, Nick McKeown, and Frode Sorensen.
\newblock {Network Tokens}.
\newblock Internet-Draft draft-yiakoumis-network-tokens-02, Internet
  Engineering Task Force, December 2020.
\newblock URL
  \url{https://datatracker.ietf.org/doc/html/draft-yiakoumis-network-tokens-02}.
\newblock Work in Progress.

\bibitem[Li et~al.(2021{\natexlab{a}})Li, Peng, Voyer, Xie, Liu, Qin, Mishra,
  Ebisawa, Previdi, and Guichard]{apn-problem-statement}
Zhenbin Li, Shuping Peng, Daniel Voyer, Chongfeng Xie, Peng Liu, Zhuangzhuang
  Qin, Gyan Mishra, Kentaro Ebisawa, Stefano Previdi, and Jim Guichard.
\newblock {Problem Statement and Use Cases of Application-aware Networking
  (APN)}.
\newblock Internet-Draft draft-li-apn-problem-statement-usecases-03, Internet
  Engineering Task Force, May 2021{\natexlab{a}}.
\newblock URL
  \url{https://datatracker.ietf.org/doc/html/draft-li-apn-problem-statement-usecases-03}.
\newblock Work in Progress.

\bibitem[Li et~al.(2021{\natexlab{b}})Li, Peng, Li, Xie, Voyer, Li, Liu, Liu,
  and Ebisawa]{apnHeaderFields}
Zhenbin Li, Shuping Peng, Cong Li, Chongfeng Xie, Daniel Voyer, Xing Li, Peng
  Liu, Chang Liu, and Kentaro Ebisawa.
\newblock {Application-aware IPv6 Networking (APN6) Encapsulation}.
\newblock Internet-Draft draft-li-6man-app-aware-ipv6-network-03, Internet
  Engineering Task Force, February 2021{\natexlab{b}}.
\newblock URL
  \url{https://datatracker.ietf.org/doc/html/draft-li-6man-app-aware-ipv6-network-03}.
\newblock Work in Progress.

\bibitem[Schinazi(2021)]{masque-protocol}
David Schinazi.
\newblock {The MASQUE Protocol}.
\newblock Internet-Draft draft-schinazi-masque-protocol-03, Internet
  Engineering Task Force, March 2021.
\newblock URL
  \url{https://datatracker.ietf.org/doc/html/draft-schinazi-masque-protocol-03}.
\newblock Work in Progress.

\bibitem[Tran and Bonaventure(2019)]{ebpfTCP}
Viet-Hoang Tran and Olivier Bonaventure.
\newblock Beyond socket options: making the linux tcp stack truly extensible.
\newblock In \emph{2019 IFIP Networking Conference (IFIP Networking)}, pages
  1--9, 2019.
\newblock \doi{10.23919/IFIPNetworking46909.2019.8999401}.

\bibitem[Spear et~al.(2016)Spear, Beyer, Cittadini, and
  Saltonstall]{beyondCorpGoogleAccessProxy}
Batz Spear, Betsy (Adrienne~Elizabeth) Beyer, Luca Cittadini, and Max
  Saltonstall.
\newblock Beyondcorp: The access proxy.
\newblock \emph{Login}, 2016.

\bibitem[Gilman and Barth(2017)]{oreillyZeroTrust}
Evan Gilman and Doug Barth.
\newblock \emph{Zero Trust Networks}.
\newblock O'Reilly Media, Inc., Sebastopol, CA, USA, 2017.
\newblock ISBN 9781491962190.

\bibitem[Gordeychik et~al.(2018)Gordeychik, Kolegov, and
  Nikolaev]{sdwanInetCensus}
Sergey Gordeychik, Denis Kolegov, and Antony Nikolaev.
\newblock Sd-wan internet census.
\newblock 2018.
\newblock URL \url{https://arxiv.org/abs/1808.09027}.

\bibitem[Mahalingam et~al.(2014)Mahalingam, Dutt, Duda, Agarwal, Kreeger,
  Sridhar, Bursell, and Wright]{rfc7348vxlan}
Mallik Mahalingam, Dinesh Dutt, Kenneth Duda, Puneet Agarwal, Larry Kreeger,
  T.~Sridhar, Mike Bursell, and Chris Wright.
\newblock {Virtual eXtensible Local Area Network (VXLAN): A Framework for
  Overlaying Virtualized Layer 2 Networks over Layer 3 Networks}.
\newblock RFC 7348, August 2014.
\newblock URL \url{https://rfc-editor.org/rfc/rfc7348.txt}.

\bibitem[Iyengar and Thomson(2021)]{rfc9000quic}
Jana Iyengar and Martin Thomson.
\newblock {QUIC: A UDP-Based Multiplexed and Secure Transport}.
\newblock RFC 9000, May 2021.
\newblock URL \url{https://rfc-editor.org/rfc/rfc9000.txt}.

\bibitem[Coninck and Bonaventure(2021)]{deconinck-quic-multipath-07}
Quentin~De Coninck and Olivier Bonaventure.
\newblock {Multipath Extensions for QUIC (MP-QUIC)}.
\newblock Internet-Draft draft-deconinck-quic-multipath-07, Internet
  Engineering Task Force, May 2021.
\newblock URL
  \url{https://datatracker.ietf.org/doc/html/draft-deconinck-quic-multipath-07}.
\newblock Work in Progress.

\bibitem[Thomson and Turner(2021)]{rfc9001quicTLS}
Martin Thomson and Sean Turner.
\newblock {Using TLS to Secure QUIC}.
\newblock RFC 9001, May 2021.
\newblock URL \url{https://rfc-editor.org/rfc/rfc9001.txt}.

\bibitem[Bosshart et~al.(2014)Bosshart, Daly, Gibb, Izzard, McKeown, Rexford,
  Schlesinger, Talayco, Vahdat, Varghese, and Walker]{p4paper}
Pat Bosshart, Dan Daly, Glen Gibb, Martin Izzard, Nick McKeown, Jennifer
  Rexford, Cole Schlesinger, Dan Talayco, Amin Vahdat, George Varghese, and
  David Walker.
\newblock P4: Programming protocol-independent packet processors.
\newblock \emph{SIGCOMM Comput. Commun. Rev.}, 44\penalty0 (3):\penalty0
  87–95, July 2014.
\newblock ISSN 0146-4833.
\newblock \doi{10.1145/2656877.2656890}.
\newblock URL \url{https://doi.org/10.1145/2656877.2656890}.

\end{thebibliography}
\end{small}

\end{document}